\documentclass[twocolumn,superscriptaddress,showpacs,preprintnumbers,amsmath,amssymb,prb]{revtex4}

\def\lesssim{\ \raise.3ex\hbox{$<$}\kern-0.8em\lower.7ex\hbox{$\sim$}\ }
\def\gesim{\ \raise.3ex\hbox{$>$}\kern-0.8em\lower.7ex\hbox{$\sim$}\ }

\usepackage{graphicx}
\usepackage{dcolumn}
\usepackage{bm}
\usepackage{amsmath,amssymb}

\begin{document}

\title{Spin imbalance effect on Larkin-Ovchinnikov-Fulde-Ferrel state}
\author{Ryosuke Yoshii}
\affiliation{Faculty of Science and Technology, Keio University, 3-14-1 Hiyoshi,
Kohoku-ku, Yokohama 223-8522, Japan}

\author{Shunji Tsuchiya}
\affiliation{Department of Physics, Faculty of Science, Tokyo University of Science, 1-3 Kagurazaka, Shinjuku-ku, Tokyo 162-8601, Japan}
\affiliation{Research and Education Center for Natural Sciences, Keio University, 4-1-1 Hiyoshi, Kanagawa 223-8521, Japan
}
\affiliation{CREST(JST), 4-1-8 Honcho, Saitama 332-0012, Japan}
\author{Giacomo Marmorini}
\affiliation{Department of Physics, Faculty of Science, Tokyo University of Science, 1-3 Kagurazaka, Shinjuku-ku, Tokyo 162-8601, Japan}
\affiliation{Research and Education Center for Natural Sciences, Keio University, 4-1-1 Hiyoshi, Kanagawa 223-8521, Japan
}
\author{Muneto Nitta}
\affiliation{Department of Physics, Keio University, 4-1-1 Hiyoshi, Kanagawa 223-8521, Japan
}
\affiliation{Research and Education Center for Natural Sciences, Keio University, 4-1-1 Hiyoshi, Kanagawa 223-8521, Japan
}

\date{\today}

\begin{abstract}
We study spin imbalance effects on 
the Larkin-Ovchinikov-Fulde-Ferrel (LOFF) state 
relevant for superconductors under a strong magnetic
field and spin polarized ultracold Fermi gas. 
We obtain the exact solution for the condensates 
with arbitrary spin imbalance 
and the fermion spectrum perturbatively in the presence of small 
spin imbalance. 
We also obtain fermion zero mode exactly 
without perturbation theory. 

\end{abstract}


\maketitle

\par
\section{Introduction}

The exotic superconducting state called 
Larkin-Ovchinikov-Fulde-Ferrel (LOFF)
state has been proposed to arise in superconductors under a strong magnetic
field.\cite{LO,FF} The LOFF state has a spatially varying order
parameter associated with Cooper pairs with finite center-of-mass momentum. 
If a strong magnetic field induces spin polarization, such Cooper pairs
are considered to form between electrons with different Fermi momenta.
The LOFF state is also relevant for the physics of color
superconductivity where quarks with different masses form pairs.\cite{Casalbuoni} 
This state was not observed for over 40 years since its
proposal, in spite of tremendous efforts.
In the last couple of years, there have been 
several claims of its observation in 
heavy fermion materials \cite{heavy} and
organic superconductors \cite{organic}, 
but direct confirmation is yet to be given
(see Ref.~\onlinecite{review} for a review).

\par
Recent developments of research in cold atomic Fermi gases have
renewed interest in the LOFF state (see Ref.~\onlinecite{Radzihovsky} for
a review). 
In two component Fermi gases consisting of atoms in two different hyperfine states, (pseudo-)spin polarization
can be controlled by changing the populations of the two components.\cite{Radzihovsky} 
Furthermore, atomic interaction can be tuned in this system by using the Feshbach
resonance which allows one to explore the interesting BCS-BEC crossover physics.
Thus, a spin polarized Fermi gas is an ideal system for realizing and
exploring the LOFF state. In Fermi gases in a toroidal trap, it has been
shown that a new state called angular LOFF state is possible in
which the rotational symmetry is spontaneously broken,\cite{Yanase} 
instead of the translational symmetry for the usual LOFF state.
Recently, observation of spin polarized superfluid state was
reported \cite{Liao} and it is expected that the LOFF state has been
achieved in this experiment. However, direct observation of its
oscillating order parameter is still lacking.
\par
The Bogoliubov-de Gennes (BdG) equation has been widely employed to
study the LOFF state. Machida and Nakanishi \cite{Machida} derived the
self-consistent LOFF state solution of the 1D BdG equation making use of
the analytical solutions of the 1D Peierls problem.\cite{BKM,Horovitz:1981zz,Mertsching,Takayama,Buzdin}
However, they assume that electrons with up and down spins have the same
Fermi velocities ($v_{{\rm F}\uparrow}=v_{{\rm F}\downarrow}$), so that
their LOFF solutions are valid only when spin polarization is small. 
This assumption is appropriate to superconducting states, 
because in ordinary superconductors the splitting of the Fermi surfaces is of 
an order of the pair potential at the Pauli limit, which is much smaller
than the radius of the Fermi surfaces.\cite{Machida}
On the other hand, the Fermi surface mismatch is in general not
small for cold atomic Fermi gases \cite{Radzihovsky} and therefore we
have to take into account large spin polarization. The spin imbalance effect was
previously studied in the Peierls problem.\cite{Brazovskii}

\par
Recently a new approach for solving the BdG equation has been proposed
by Ba\c{s}ar and Dunne.\cite{Basar} They derived 
the non-linear Schr\"odinger equation (NLSE) 
for the order parameter $\Delta$
with a suitable ansatz for Gor'kov Green's function.
Since the derived NLSE is a
closed equation for the order parameter $\Delta(x)$, this enables one to 
avoid the self-consistent calculation of the coupled
equations of the BdG equation and the gap equation.
Using this approach, they found a new self-consistent solution for a complex kink
crystal, which includes all previously known solutions as special cases,
such as the solutions of the LOFF state 
(real kink crystal) \cite{Machida} and 
Shei's complex (twisted) kink \cite{Shei}. 
This new approach and the complex kink crystal solution have been
further developed \cite{Basar} for the massless Gross-Neveu model
\cite{Gross:1974jv} and the Nambu-Jona-Lasinio model in $1+1$ dimension. \cite{Nambu:1961tp} 
However, this approach has not been extended to spin polarized system.
\par
In this paper, we investigate the self-consistent solutions of the BdG
equation for spin {\it imbalanced} Fermi 
condensates. 
We extend the approach developed by Ba\c{s}ar and Dunne to obtain the
analytic solutions for the LOFF state in which the order parameter
exhibits spatial oscillations. 
In contrast to the solutions of Machida and Nakanishi, we take into account the
difference in the Fermi velocities ($v_{{\rm F}\uparrow} \neq v_{{\rm
F}\downarrow}$) and derive the exact solutions for the condensate $\Delta(x)$ which are valid for {\it any} spin polarizations.
We show how the effect of large spin polarization changes the form of
the non-linear Schr\"odinger equation for the order parameter.
We also develop the perturbation theory for the BdG equation in the
presence of small spin polarization, and 
obtain the fermion zero mode which is exact for arbitrary spin polarization.

\section{Nonlinear Schr\"odinger equation for order parameter}
In this section, we derive the nonlinear Schr\"odinger equation for the
order parameter $\Delta$ in the presence of spin polarization.

\subsection{The Bogolibov-de Gennes equation}
We consider a gas of fermions with spin up and down in
quasi-one dimension under a magnetic field. If fermions with
different spins interact attractively, the system undergoes a 
superconducting (superfluid) transition at low temperature.  
Although the mean field approximation is not valid in strict one dimension, 
since we assume a quasi-one dimensional system relevant for experiments, 
the system can be described by the mean-field BdG equation
\cite{deGennes} 
(we set $\hbar=1$)
\begin{eqnarray}
&\left[
\begin{array}{cc}
H_{\uparrow}(x)&\Delta_0(x)\\ 
\Delta_0^\ast(x) &-H_{\downarrow}(x)
\end{array}
\right]
\left[
\begin{array}{c}
u_0(x) \\ v_0(x)
\end{array}
\right]
=E
\left[
\begin{array}{c}
u_0(x) \\ v_0(x)
\end{array}
\right],\\
& \displaystyle H_{\sigma}(x)=-\frac{1}{2M}\frac{\partial^2}{\partial x^2}-\mu_\sigma,
\end{eqnarray}
where $\sigma$($=\uparrow, \downarrow$) stands for the spin and $M$ is
the mass of the fermion. 
The energy difference due to the Zeeman splitting is included in
the chemical potential for each spin state $\mu_\sigma$
($\sigma=\uparrow,\downarrow$). This model is indeed applicable to an
imbalanced cold Fermi gas.\cite{Radzihovsky} 
Throughout this paper, we restrict ourselves at $T=0$. 
In this case, the order parameter $\Delta_0(x)$ satisfies the gap equation
\begin{equation}
\Delta_0(x)=-2g^2\sum_{E_n<0}u_n(x)v_n(x)^*,
\end{equation}
where $g$ is the attractive interaction between fermions with different
spins and $n$ is the index for eigenstates. 

If the attractive interaction is small compared with the Fermi energy
$\varepsilon_{{\rm F}\sigma}= \mu_\sigma$, fermions near the Fermi
surfaces form Cooper pairs. If we assume
$u_0(x)=e^{ik_{\mathrm{F}\uparrow}x} u(x)$ and 
$v_0(x)=e^{-ik_{\mathrm{F}\downarrow}x} v(x)$ ($k_{\rm F\sigma}$
is the Fermi momentum $k_{\rm F\sigma}=\sqrt{2M\varepsilon_{\rm F\sigma}}$),
$u(x)$ and $v(x)$ vary much slower than $1/k_{\mathrm{F}\sigma}$.
Neglecting the second derivative term of $u(x)$ and $v(x)$ (the Andreev
approximation \cite{Barsagi}), the BdG
equation reduces to 
\begin{eqnarray}
&\left[
\begin{array}{cc}
-i v_{\mathrm{F}\uparrow}\frac{\partial}{\partial x}&\Delta(x)\\ 
\Delta^\ast(x) & i v_{\mathrm{F}\downarrow}\frac{\partial}{\partial x}
\end{array}
\right]
\left[
\begin{array}{c}
u(x) \\ v(x)
\end{array}
\right]
=E
\left[
\begin{array}{c}
u(x) \\ v(x)
\end{array}
\right],
\label{BdG}
\end{eqnarray}
where $v_{\mathrm{F}\sigma}=k_{\mathrm{F}\sigma}/m$ is the Fermi velocity and
$\Delta=e^{-i(k_{\mathrm{F}\uparrow}+k_{\mathrm{F}\downarrow})x}\Delta_0$.
When $v_{\mathrm{F}\uparrow}=v_{\mathrm{F}\downarrow}$, the LOFF state
solution of Eq.~(\ref{BdG}) has been derived in Refs.~\onlinecite{Machida,Basar}.

\subsection{The Ba\c{s}ar-Dunne formalism}

In Ref.~\onlinecite{Basar}, the so-called nonlinear
Schr\"odinger equation (NLSE) for $\Delta(x)$ has been derived for the
case of $v_{\mathrm{F}\uparrow}=v_{\mathrm{F}\downarrow}$, 
through the analysis of Gor'kov Green's function. This is a
convenient way to solve the BdG equation. Since the derived NLSE is a
closed equation for $\Delta(x)$, this enables one to 
avoid the self-consistent calculation of the coupled
equations of the BdG equation and the gap equation.
We extend this analysis to the case of $v_{\mathrm{F}\uparrow}\neq
v_{\mathrm{F}\downarrow}$.
 
First, we derive the Gor'kov Green's function that satisfies
\begin{equation}
(H-E)G(x,y;E)=\delta(x-y),
\label{defGf}
\end{equation}
where
\begin{eqnarray}
H=\left[
\begin{array}{cc}
-i v_{\mathrm{F}\uparrow}\frac{\partial}{\partial x}&\Delta(x)\\ 
\Delta^\ast(x) &i v_{\mathrm{F}\downarrow}\frac{\partial}{\partial x}
\end{array}
\right]. 
\end{eqnarray}
The Gor'kov Green's function can be constructed from two independent
solutions $\psi(x)$ and $\phi(x)$ of Eq.~(\ref{BdG}) as \cite{Kosztin} 
\begin{eqnarray}
G(x,y;E)&=\left(
\begin{array}{cc}
0 & v_{\mathrm{F}\uparrow}^{-1}\\
v_{\mathrm{F}\downarrow}^{-1} & 0 
\end{array}
\right)F^\ast(x,y;E),\label{Gf}
\\
F(x,y;E)&=
\frac{1}{iW(x)}\left[\theta(y-x)\psi(x)\phi^T(y)\right.\nonumber\\ 
&\left.+\theta(x-y)\phi(x)\psi^T(y)\right],
\end{eqnarray}
where $W\equiv i\psi^T\sigma_2\phi$ is a Wronskian. 
It is easy to show that the Eq.\ (\ref{Gf}) satisfies (\ref{defGf}). 

On the other hand, the diagonal resolvent is defined by
\begin{equation}
R(x;E)=
\left< x\left|\frac{1}{H-E}\right|x \right>
 \label{defresol}.
\end{equation}
Indeed, Eq.~(\ref{defresol}) includes all spectral information for fermions
in the presence of $\Delta(x)$, such as the single-particle spectral function
\begin{equation}
\rho(E)=\frac{1}{\pi}\mathrm{Im}\int dx\mathrm{Tr}R(x;E+i\delta).
\label{spectrum-fn}
\end{equation}

From Eq.~(\ref{Gf}), $R(x;E)$ can be obtained as the coincident
limit of Gor'kov Green's function: 
\begin{eqnarray}
R(x;E)
=\lim_{\delta\to 0^+} \frac{1}{2}[G(x,x+\delta ;E)+G(x+\delta ,x;E)].
\label{resol}
\end{eqnarray}
We can show that $R(x)$ satisfies the following conditions:
\begin{eqnarray}
&\mathrm{Tr}\tilde R(x;E)\sigma_3=0,\label{traceless}\\  
&\det \tilde R(x;E)
=-\frac{1}{4}\label{det},
\end{eqnarray}
where 
\begin{eqnarray}
\tilde R(x;E)\equiv\left(
\begin{array}{cc}
v_{\mathrm{F}\uparrow} & 0\\
0 & v_{\mathrm{F}\downarrow}
\end{array}
\right)R(x;E).
\end{eqnarray}
In addition to the above conditions, from the definition
Eq.~(\ref{defresol}), the resolvent must satisfy the Hermiticity condition:
\begin{equation}
R^\dagger(x;E)=R(x;E).
\end{equation}
By a straight forward calculation, we obtain 
\begin{align}
&\partial_x \tilde R(x;E)\sigma_3\nonumber =\\
& i\left[\left(
\begin{array}{cc}
v_{\mathrm{F}\downarrow}^{-1}&0\\ 
0&v_{\mathrm{F}\uparrow}^{-1}
\end{array}
\right)
\left(
\begin{array}{cc}
E&-\Delta(x)\\ \Delta^\ast(x) &-E
\end{array}
\right),\tilde R(x;E)\sigma_3\right].\label{eilen}
\end{align}
Eq.~(\ref{eilen}) is known as the Dikii-Eilenberger equation. \cite{Kosztin}

Note that we have derived the above equation from the BdG equation
only with the Andreev approximation. On the other hand, it is well known
that the Eilenberger equation can be derived from the BdG equation or equivalently
Gor'kov equation using the quasiclassical approximation in addition to
the Andreev approximation in 3D.\cite{Ichioka}
In the present case, since we assume the system is (quasi) 1D where the Andreev
approximation and the quasiclassical approximation are equivalent, the
Dikii-Eilenberger equation can be derived without explicitly using the
quasiclassical approximation. Therefore, the solutions of
Eq.~(\ref{eilen}) are the exact self-consistent solutions of the BdG equation.

Next step is to make an ansatz for the form of the Gor'kov Green's function. 
From the gap equation, $\Delta(x)$ must satisfy 
\begin{equation}
\Delta(x)
\propto \frac{\delta}{\delta\Delta(x)^\ast}\int dE \rho(E)
\ln\left(1+e^{-\beta(E-\mu)}\right).
\label{gap1}
\end{equation}
From Eq.~(\ref{spectrum-fn}), the simplest ansatz for $R(x;E)$ to satisfy Eq.\ (\ref{gap1}) 
is that the diagonal entries are set to be proportional to $|\Delta(x)|^2$. 
The gap equation can be derived by the functional derivative  
\begin{equation}
\Delta(x)\propto \mathrm{Tr}_{D,E}\left[\sigma_1(1+\sigma_3)R(x,E)\right].
\label{gap2}
\end{equation}
The simplest solution for Eq.\ (\ref{gap2}) is for off-diagonal entries of the 
$R(x;E)$ to be proportional to $\Delta(x)$ (or $\Delta^\ast(x)$). 
However, the consistency between (\ref{eilen}) and (\ref{gap2}) requires that 
the derivative term $\Delta^\prime(x)$ (or $\Delta^{\ast\prime}(x)$) should be in 
the off-diagonal entries. 
The last condition comes from Eq.\ (\ref{traceless}). 
If we assume the following form for the resolvent
\begin{align}
& R(x;E)=\nonumber\\
& \mathcal{N}\left(
\begin{array}{cc}
v_{\mathrm{F}\downarrow}(a+|\Delta(x)|^2)&
b\Delta(x)-ic\Delta^\prime(x)\\
b\Delta^\ast(x)+ic\Delta^{\ast\prime}(x)&
v_{\mathrm{F}\uparrow}(a+|\Delta(x)|^2)
\end{array}
\right),
\label{ansatz}
\end{align}
by substituting Eq.~(\ref{ansatz}) into the right hand side of
Eq.~(\ref{eilen}), we obtain
\begin{equation}
\partial_x \tilde R(x;E)\sigma_3=\mathcal{N}(E)\left(
\begin{array}{cc}
A&B
\\
-B^\ast&-A
\end{array}
\right),
\end{equation} 
where 
\begin{eqnarray}
A&=&c(|{\Delta}|^2)^\prime,\\
B&=&iv_{\mathrm{F}\downarrow}(v_{\mathrm{F}\uparrow}^{-1}+v_{\mathrm{F}\downarrow}^{-1})
E(b(E)\Delta^\ast+ic\Delta^{\ast\prime})\nonumber\\
&&-2iv_{\mathrm{F}\downarrow}\Delta^\ast 
(a(E)+|\Delta|^2).
\end{eqnarray}
Then, we find
\begin{equation}
c=v_{\mathrm{F}\uparrow}v_{\mathrm{F}\downarrow},
\end{equation}
from the diagonal part, and 
\begin{equation}
\tilde{\Delta}^{\prime\prime}+i
[\tilde{b}-2\tilde E]\tilde{\Delta}^{\prime}
-2[\tilde{a}-\tilde{E}\tilde{b}]\tilde\Delta 
-2\tilde\Delta|\tilde{\Delta}|^2=0,
\label{NLSEim}
\end{equation}
from the off diagonal part. 
Here, we have defined
\begin{equation}
\tilde a=\alpha^{-2} a,
\ 
\tilde b=\alpha^{-2} b,
\ 
\tilde E=\alpha^{-2}E,
\ 
\tilde \Delta=\alpha^{-1}\Delta,
\label{tilde}
\end{equation}
where $\alpha$ is the 
imbalance parameter, that is
\begin{equation}
\alpha\equiv\frac{2\sqrt{v_{\mathrm{F}\uparrow}v_{\mathrm{F}\downarrow}}}
{v_{\mathrm{F}\uparrow}+v_{\mathrm{F}\downarrow}}=
\frac{\sqrt{v_{\mathrm{F}\uparrow}v_{\mathrm{F}\downarrow}}}
{v_{\mathrm{F}}}.
\label{scaling}
\end{equation}
$0\leq\alpha\leq1$. We introduce the 
mean Fermi velocity by
$v_{\mathrm{F}}=(v_{\mathrm{F}\uparrow}+v_{\mathrm{F}\downarrow})/2$,
and set $v_{\mathrm{F}}=1$.
In the balanced case ($v_{\rm F\uparrow}=v_{\rm F\downarrow}$), it has
been shown that this equation reproduce the well known solutions, e.g., homogeneous condensate (BCS theory), 
single kink condensate \cite{Takayama}, and real kink crystal \cite{Basar}.

To confirm the consistency condition Eq.~(\ref{det}), we calculate the determinant of the resolvent:
\begin{eqnarray}
\det \tilde R&=&\alpha^8\mathcal{N}^2 \left[
|\tilde\Delta|^4-|\tilde\Delta^\prime|^2+(2\tilde a-\tilde b^2)|\tilde\Delta|^2\right.\nonumber\\
&&\left.+i\tilde b({\tilde\Delta}^\prime{\tilde\Delta}^\ast-{\tilde\Delta}{\tilde\Delta}^{\ast\prime})+\tilde a^2
\right].
\end{eqnarray}
The NLSE implies that the right hand side of the above equation is constant as follows
\begin{eqnarray}
\frac{d}{dx}\left(\frac{\det \tilde R}{\alpha^8\mathcal{N}^2}\right)&=&(
2|\tilde\Delta|^2+2\tilde a-\tilde b^2)(|\tilde\Delta|^2)^\prime\nonumber\\
&&-({\tilde\Delta}^{\prime\prime}{\tilde\Delta}^{\ast\prime}
+{\tilde\Delta}^\prime{\tilde\Delta}^{\ast\prime\prime})\nonumber\\
&&+i\tilde b(\tilde\Delta^{\prime\prime}\tilde\Delta^\ast
-\tilde\Delta^{\ast\prime\prime}\tilde\Delta)=0.
\end{eqnarray}

It is remarkable that by the scalings in Eq.~(\ref{tilde}), the NLSE (\ref{NLSEim})
for finite spin polarization takes exactly the same form as the one for
zero spin polarization. This means that there exist solutions of Eq.~(\ref{NLSEim})
which correspond to each of the solutions of the NLSE for the balanced case. 
Thus, the exact solutions of Eq.~(\ref{NLSEim}) can be easily derived by scaling the
solutions of the NLSE for the balanced case.
The solution corresponding to a complex kink crystal is the most general one
which includes other solutions in some limits.
We derive the solutions of Eq.~(\ref{NLSEim}) in Sec.~\ref{sec4} including that
of the LOFF state.

\section{Imbalance effect on single-particle states}

In this section, we examine the effect of spin imbalance on
single-particle states. For simplicity, we consider the case when the
order parameter is real. 

\subsection{Imbalance effect on fermionic zero mode}

First, we consider the fermionic zero mode, i.e. the solution of
Eq.~(\ref{BdG}) with $E=0$.
The zero mode plays crucial roles for the LOFF state. The wave
function of the zero mode is localized around the nodes of the order
parameter and accommodate the excess spin component.
For the zero mode solution, we can exactly solve the BdG equation.
When $E=0$, by the scaling transformations
\begin{eqnarray}
\tilde u\equiv{(1+\epsilon/2)}^{-\frac{1}{2}}\,u, \ \ 
\tilde v\equiv{(1-\epsilon/2)}^{-\frac{1}{2}}\,v, \label{transuv}\\
\tilde x\equiv (1-\epsilon^2/4)x=\alpha^{2}x,\label{transx}
\end{eqnarray}  
Eq.~(\ref{BdG}) can be rewritten as
\begin{eqnarray}
&\left[
\begin{array}{cc}
-i\frac{\partial}{\partial {\tilde x}}&\tilde\Delta\\ 
\tilde\Delta &i\frac{\partial}{\partial {\tilde x}}
\end{array}
\right]
\left[
\begin{array}{c}
\tilde u \\ \tilde v
\end{array}
\right]
=0,
\label{eq:zeromode}
\end{eqnarray}
where $\epsilon\equiv v_{\mathrm{F}\uparrow}-v_{\mathrm{F}\downarrow}$.
It is clear that the above equation has the same form as the one for $v_{\rm
F\uparrow}=v_{\rm F\downarrow}$. This indicates that if the BdG equation
has a zero mode solution for balanced case, there exists a
corresponding zero mode solution for imbalanced case, and the two
solutions are related by the scaling transformations (\ref{transuv})
and (\ref{transx}).

Furthermore, the zero mode solution can be explicitly constructed as
follows. If one applies the unitary transformation 
\begin{equation}
\left[
\begin{array}{c}
\tilde f_+\\
\tilde f_-
\end{array}
\right]
=
\frac{1}{\sqrt{2}}
\left[
\begin{array}{cc}
1 & -i \\
i & -1
\end{array}
\right]
\left[
\begin{array}{c}
\tilde u\\
\tilde v
\end{array}
\right],
\end{equation} 
Eq.\ (\ref{eq:zeromode}) yields 
\begin{equation}
\left[\partial_{\tilde x}\mp\tilde\Delta\right]\tilde f_\pm =0.
\label{zerostateeq}
\end{equation}
Hence, the solution of Eq.\ (\ref{zerostateeq}) can be formally written as 
\begin{equation}
\tilde f_\pm (x)\propto \exp\left[{\pm\int^x_0dy\ \alpha^2\tilde\Delta(y)}\right]. 
\label{zerostate}
\end{equation}
Equation~(\ref{zerostate}) is valid if $\tilde f_\pm$ is
normalizable. That is $\int_{-\infty}^{\infty} dx\tilde f_\pm^2$ is finite.
The solution (\ref{zerostate}) is exactly the same 
as the one for the balanced case up to the scaling factor for the order parameter. 

\subsection{Perturbation theory for massive modes}

We develop a perturbation theory for massive modes
($E>0$). We calculate spin imbalance correction for the solutions of Eq.\ (\ref{BdG}) 
perturbatively taking $\epsilon$ as a
small parameter. We expand the solution of Eq.~(\ref{BdG}) by $\epsilon$ as
\begin{eqnarray}
\left[
\begin{array}{c}
u(x) \\ v(x)
\end{array}
\right]
&=&
\left[
\begin{array}{c}
u^{(0)}(x) \\ v^{(0)}(x)
\end{array}
\right]+
\epsilon\left[
\begin{array}{c}
u^{(1)}(x) \\ v^{(1)}(x)
\end{array}
\right]+O(\epsilon^2),
\label{BdGP}\\
E&=&E^{(0)}+\epsilon E^{(1)}+O(\epsilon^2).
\label{EP}
\end{eqnarray}

In the 0th order, we indeed obtain the equation for balanced case
\begin{eqnarray}
&\left[
\begin{array}{cc}
-i\frac{\partial}{\partial x}&\tilde\Delta(x)\\ 
\tilde\Delta(x) &i\frac{\partial}{\partial x}
\end{array}
\right]
\left[
\begin{array}{c}
 u^{(0)}(x) \\  v^{(0)}(x)
\end{array}
\right]
=E^{(0)}
\left[
\begin{array}{c}
 u^{(0)}(x) \\ v^{(0)}(x)
\end{array}
\right].
\label{0th}
\end{eqnarray}
Here, we used
$\Delta=\alpha\tilde\Delta=
\sqrt{1-\epsilon^2/4}\tilde\Delta\approx(1-\epsilon^2/8)\tilde\Delta$.
Making use of the unitary transformation 
\begin{equation}
\left[
\begin{array}{c}
f_+\\
f_-
\end{array}
\right]
=
\frac{1}{\sqrt{2}}
\left[
\begin{array}{cc}
1 & -i \\
i & -1
\end{array}
\right]
\left[
\begin{array}{c}
 u\\
 v
\end{array}
\right],
\end{equation} 
Eq.~(\ref{0th}) becomes
\begin{equation}
\left(
-\partial_x^2\mp\tilde\Delta^\prime+\tilde\Delta^2-E^{(0)}
\right)
f^{(0)}_{\pm}(x)=0.
\label{f0th}
\end{equation}
Once $\tilde \Delta$ is obtained by solving Eq.~(\ref{NLSEim}),
the Shr\"odinger type equation (\ref{f0th}) yields a set of unperturbed
eigenstates.

From the first order terms, we obtain
\begin{eqnarray}
&\left(
-\partial_x^2\mp\tilde\Delta^\prime+\tilde\Delta^2-E^{(0)}
\right)
f^{(1)}_{\pm}(x)
+\frac{i}{2}\tilde\Delta f^{(0)}_{\mp}(x)\nonumber\\
&+iE^{(0)}\partial_x f^{(0)}_{\pm}(x)
-2E^{(0)}E^{(1)}f^{(0)}_{\pm}(x)=0.
\label{eq:f}
\end{eqnarray}
Thus, the first order correction for the energy can be calculated 
from the nonperturbative eigenvalues $E_n^{(0)}$ and eigenstates
$f_{\pm,n}^{(0)}$ as 
\begin{eqnarray}
E^{(1)}_n=
-\frac{i}{2}\int dx 
\left[
f_{+,n}^{(0)\ast}(x) \ f^{(0)\ast}_{-,n}(x)
\right]
\partial_{x}
\left[
\begin{array}{c}
f^{(0)}_{+,n}(x) \\ f^{(0)}_{-,n}(x)
\end{array}
\right].
\label{1st}
\end{eqnarray}

\section{Spin imbalance correction for various condensates}\label{sec4} 

\subsection{Homogeneous condensate}
Here we consider the homogeneous condensate 
\begin{equation}
\tilde\Delta(x)=m.
\label{hc}
\end{equation}
We can always take $m$ to be real due to the chiral symmetry of the model. 
The substitution Eq.\ (\ref{hc}) into Eq.\ (\ref{NLSEim}) yields \cite{caution1} 
\begin{equation}
\tilde a=2\tilde E^2 -m^2,\ \tilde b=2\tilde E.
\label{hcres}
\end{equation}

Next, we calculate the energy spectrum for the quasiparticles. 
In this case, the order parameter is constant and thus the zero mode 
(\ref{zerostate}) is not allowed, {\it i.e.} it is not normalizable. 
Then we calculate the massive modes. 
From Eq.\ (\ref{0th}), we obtain 
\begin{eqnarray}
&\left[
\begin{array}{cc}
-i\frac{\partial}{\partial x}&m\\ 
m &i\frac{\partial}{\partial x}
\end{array}
\right]
\left[
\begin{array}{c}
u^{(0)}(x) \\ v^{(0)}(x)
\end{array}
\right]
=E^{(0)}
\left[
\begin{array}{c} 
u^{(0)}(x) \\ v^{(0)}(x)
\end{array}
\right].
\end{eqnarray}
Then we obtain 
\begin{equation}
E(k)_\pm^{(0)}=\pm\sqrt{k^2+m^2},
\end{equation}
and the eigenspinor is  
\begin{eqnarray}
\left[
\begin{array}{c}
 u^{\pm(0)}_k(x) \\ v^{\pm(0)}_k(x)
\end{array}
\right]=e^{ikx}
\left[
\begin{array}{c}
 u_k^\pm \\ v_k^\pm
\end{array}
\right],
\label{h0th}
\end{eqnarray}
where $u_k$ and $v_k$ is independent of $x$.
Substituting Eq.\ (\ref{h0th}) into Eq.\ (\ref{1st}) yields
\begin{equation}
E(k)_\pm^{(1)}=\frac{1}{2}k.
\end{equation}
Then we obtain the energy dispersion 
\begin{equation}
E(k)_\pm=\pm\sqrt{k^2+m^2}+\frac{\epsilon}{2}k+O(\epsilon^2).
\label{correction-homo}
\end{equation}

\begin{center}
\begin{figure}[h]
\includegraphics[width=0.3\textwidth]{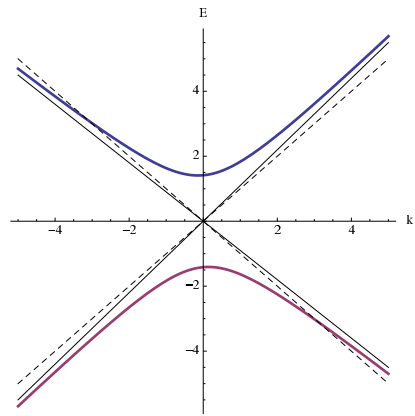}
\caption{Fermionic spectrum in the case of homogeneous condensate
 $\Delta=2$ with $\epsilon=0.2$. The slope of the asymptotes changes by
 $\epsilon/2$ as a consequence of the spin imbalance. The dashed line
 shows the asymtotes for the spectrum when $\epsilon=0$.}
\label{spectrum}
\end{figure}
\end{center}

In this case, the spectrum of the BdG equation (\ref{BdG}) can also be calculated exactly. 
By substituting Eq.\ (\ref{tilde}) and Eq.\ (\ref{hc}) into Eq.\ (\ref{BdG}), we obtain
\begin{equation}
E_\pm(k)=\pm\sqrt{k^2+\alpha^2m^2}+\frac{\epsilon}{2} k.
\end{equation}
This dispersion relation is plotted in Fig.~\ref{spectrum}. When $\epsilon\ll 1$, $E\simeq \frac{\epsilon}{2} k\pm\sqrt{k^2+m^2}$, which is consistent with the result (\ref{correction-homo}) 
obtained by the perturbation theory.

When $|k|\gg\alpha^2m^2$, $E\simeq \pm |k|+\epsilon k/2$. 
This indicates that the energy dispersion asymptotically becomes that of 
the free fermion ($E=v_{\mathrm{F}\uparrow}k, -v_{\mathrm{F}\downarrow}k$).

Note that the energy gap contracts by a factor $\alpha^2$ compared to the balanced case, namely the edges of the positive and negative energy bands become $\pm m\alpha^2$
as plotted in Fig.\ \ref{spectrum}.

\subsection{Single real kink condensate}
Next solution we consider is the single real kink (anti-kink) condensate
\begin{equation}
\tilde\Delta(x)=\pm m\tanh (mx).
\label{srk}
\end{equation}
This solution can be obtained by setting 
\begin{equation}
\tilde a=2\tilde E^2-m^2,\ \tilde b=2\tilde E. 
\end{equation}
Now we analyze the spectrum of the associated BdG equation. 
The energy spectrum for single kink condensate is obtained 
in the limit of infinite 
periodicity of the real kink crystal, which will be discussed in the following section. 
The only exception is that the normalizable zero mode exists in this case. 
For the real kink case (anti-kink case), the eigenstate 
$\tilde f_+(x)$ ($\tilde f_-(x)$) in Eq.\ (\ref{zerostate}) is normalizable 
and $\tilde f(x)$ ($=\tilde f_+(x)$ for kink, $=\tilde f_-(x)$ for anti-kink) 
becomes
\begin{eqnarray}
\tilde f(x)= N\left[\mathrm{sech}(mx)\right]^{\alpha^2}, 
\end{eqnarray}
where $N$ is the normalization constant.  

\subsection{LOFF state}

As shown in Ref.~\onlinecite{Basar}, the NLSE for balanced case has the LOFF
solution (real kink crystal). 
Then, we can immediately conclude that Eq.~(\ref{NLSEim}) has 
the corresponding solution
\begin{equation}
\tilde\Delta(x)=\sqrt{\nu}\frac{2m}{1+\sqrt{\nu}}\mathrm{sn}
\left(
\frac{2m}{1+\sqrt{\nu}}x;\nu
\right),
\label{rk}
\end{equation}
where $\mathrm{sn}$ is the Jacobi elliptic function with 
real elliptic parameter $0\le\nu\le 1$. 
Then we can conclude that the spin imbalance results in the dilatation with a factor $\alpha^2$ of the condensate. 
The substitution Eq.\ (\ref{rk}) into Eq.\ (\ref{NLSEim}) yields 
\begin{eqnarray}
\tilde a(\tilde E)&=&2\tilde E^2-2m^2\frac{1+\nu}{(1+\sqrt{\nu})^2},\\
\tilde b(\tilde E)&=&2\tilde E.
\end{eqnarray}

The eigenstates of the quasiparticles 
for the real kink crystal order parameter (\ref{rk}) are given 
as follows \cite{Machida},
\begin{eqnarray}
f_{+,n}^{(0)}(x)=&&\left[
\frac{\wp\left(x+\omega_3\right)-e}{2L\left(\bar \wp-e\right)}
\right]^{\frac{1}{2}}\nonumber\\
&& \times \exp\left[
iC\left(E\right)\int^x_0
\frac{dx^\prime}{\wp\left(x^\prime+\omega_3\right)-e}
\right], 
\label{eigen}
\end{eqnarray}
where $L$ is the size of the system and $\wp$ is the Weierstrass function which obeys 
$\tilde\Delta^2(x)-\tilde\Delta^\prime(x)=e_1+2\wp(x+\omega_3)$  
with 
\begin{eqnarray}
e_1&=&\frac{2m^2}{3(1+\sqrt{\nu})^2}(1+\nu),\\
e_2&=-&\frac{m^2}{3(1+\sqrt{\nu})^2}(1-6\sqrt{\nu}+\nu),\\
e_3&=-&\frac{m^2}{3(1+\sqrt{\nu})^2}(1+6\sqrt{\nu}+\nu),\\
e&=&e_1-E^2.
\end{eqnarray}
The amplitude of $f_{+,n}$ has the half periodicity of 
\begin{equation}
\omega=K(\nu)/m, 
\end{equation}
where $K(\nu)$ is the complete elliptic integral of the first kind. 
We define $\bar\wp$ as the average of $\wp$ 
\begin{eqnarray}
\bar\wp&=&\frac{1}{\omega}\int^\omega_0\wp(x+\omega_3)dx.
\end{eqnarray} 
The coefficient $C(E)$ in Eq.~(\ref{eigen}) is defined by 
\begin{eqnarray}
C(E)&=&\pm E\sqrt{(E^2-E_2^2)(E^2-E_3^2)},\\
E_i^2&=&e_1-e_i  \;\;(i=2,3).
\end{eqnarray}

Substituting Eq.\ (\ref{eigen}) into Eq.\ (\ref{1st}), we obtain
\begin{equation}
E^{(1)}=-\frac{C(E)}{2\left(\bar \wp-e\right)}.
\end{equation}
{This result implies, in particular, that the two gaps shrink as a consequence of non-zero imbalance, as illustrated in Fig.\ 2.}

\begin{figure}[ht]
\begin{center}
\includegraphics[width=0.3 \textwidth]{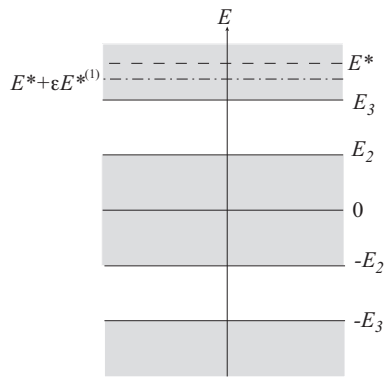}
\caption{Schematic illustration of the first order correction to the fermionic energy levels with respect to the balanced case for the real kink crystal condensate.}
\end{center}
\label{bands}
\end{figure}

In the limit of $\nu\rightarrow 1$, the periodicity becomes infinite and then 
the LOFF state (the real kink crystal condensate) reduces
to the single kink condensate.

\section{Conclusions}
We conclude this paper with few remarks. 
In this paper, we have investigated 
the spin imbalance correction for the BdG equation. 
We have expanded the method in Ref.~\onlinecite{Basar}, which is valid for the balanced case, and have obtained the non-linear 
Schr\"odinger equation for the order parameter with spin imbalance. 
We have shown that the imbalance correction for the order parameter is only 
included in the reparameterization, however this result is nontrivial without 
using this method.
We have obtained the fermionic zero-mode exactly for arbitrary spin imbalance
by a scale transformation of the one in the balanced case, 
which implies 
the stability of the fermionic zero-mode 
against the spin imbalance. 
We also have analyzed the massive fermionic spectrum 
of the BdG equation with small spin imbalance by 
the perturbation theory.  
We have applied the method for homogeneous condensate, the single kink condensate and the LOFF state (the real kink crystal condensate). 
For the homogeneous condensate, we show the con-
sistency between the perturbation theory and the exact solution for
fermionic spectrum in first order. 
For the real kink crystal condensate, we have obtained the imbalance correction for the order parameter and the fermionic spectrum. 
This result completely generalizes those of Ref.~\onlinecite{Machida}. 

Finally, we make few remarks, (i) we have analyzed the fermionic problem by 
the perturbation at the first order in the spin imbalance parameter $\epsilon$, 
however the higher expansion is straightforward. (ii) As in the 
case of homogeneous condensate, we may obtain the exact solution. 
(iii) We have dealt with the real condensate in this paper, 
however the method used here can be generalized to the complex case, 
such as the twisted kink \cite{Shei} and the twisted kink crystal \cite{Basar}.

\par

\section*{Acknowledgments}
We thank G.~Dunne, A.~Flachi, K.~Machida, M.~Ruggieri and Y.~Yanase for useful discussions. The work of R.Y. is supported by Global COE
Program ``High-Level Global Cooperation for
Leading-Edge Platform on Access Space (C12).'' 
The work of G.M.\ is supported by Japan Society for Promotion of Science. 
The work of M.N.\ is partially supported by a Grant-in-Aid for Scientific Research (Nos.\ 20740141 and 23740198) and by the "Topological
Quantum Phenomena" (No.\ 23103515) Grant-in Aid for Scientific Research on Innovative Areas from the Ministry of Education, Culture,
Sports, Science and Technology (MEXT) of Japan.


\begin{thebibliography}{}
\bibitem{caution1}
In order to obtain Eq.\ (\ref{hcres}), we put $m$ to be $me^{-\epsilon x}$ and then 
we take the limit of $\epsilon\rightarrow 0$.
\end{thebibliography}
\end{document}